\documentstyle[prb,aps,epsfig]{revtex}

\newcommand{\stk}[2]{\mbox{\begin{tabular}{l}{#1}\\{#2}\end{tabular}}}

\begin{document}
\draft
\input{epsf}
\twocolumn[\hsize\textwidth\columnwidth\hsize\csname@twocolumnfalse\endcsname

\title{Conventional and unconventional orderings in the jarosites}

\author{A.~S. Wills\cite{ASW}}

\address{D\'epartement de Recherche Fondamentale sur la Mati\`ere Condens\'ee,
SPSMS, CEA Grenoble, 38054 Grenoble, France.}

\date{\today}
\maketitle

\begin{abstract}

The jarosites make up the most studied family of {\it kagom\'e} antiferromagnets.
The flexibility of the
structure to substitution of the A and B ions allows a wide range of
compositions to be synthesised
with the general formula AB$_3$(SO$_4$)$_2$(OH)$_6$ (A = Na$^+$, K$^+$,
Ag$^+$, Rb$^+$,
H$_3$O$^+$, NH$_4^+$, $\frac{1}{2}$Ba$^{2+}$, and $\frac{1}{2}$Pb$^{2+}$;
B = Fe$^{3+}$,
Cr$^{3+}$, and V$^{3+}$). Additional chemical tuning of the exchange between
layers is also
possible by substitution of the (SO$_4$)$^{2-}$ groups by (SeO$_4$)$^{2-}$
or (CrO$_4$)$^{2-}$.
Thus, a variety of $S$ = 5/2, 3/2, and 1 systems can be engineered to allow
study of the effects of
frustration in both the classical and more quantum limits. Within this family both 
conventional 
long-ranged magnetic order and more
exotic unconventional
orderings have been found. This article reviews the  different types of
magnetic orderings that occur
and examines some of the parameters that are their cause.

\end{abstract}
\vspace{.2in} 


\noindent 
] 
\narrowtext 

\section{Introduction.}

Research in frustrated magnetic systems follows two main
directions. In the first, the effects of frustration are studied
in systems that show otherwise conventional behaviour. Rather than
being simply a study of the `soft modes' that are characteristic
of frustrated magnetism, these investigations can reveal
fundamental differences in magnetic properties and their causes.
The other, and frequently more pursued,  direction is the search
for, and study of, new and unconventional types of magnetic ground
states. Particular effort in the study of frustrated magnetism is
made on systems where the magnetic ions make up the {\it kagom\'e}
and {\it pyrochlore} geometries. This is because the frustration
of their triangular motifs is amplified by the vertex sharing
topology, and is at such a degree that the conventional N\'eel
-type of long-range order is not expected to occur even at $T=0$
in the $S=\frac{1}{2}$ nearest-neighbour systems.

In this article a review is made of the observed magnetic properties
of different members of the jarosite
family. These present examples of systems in which the {\it kagom\'e} lattice
of magnetic ions can be
decorated by ions with a variety of spin states. They therefore provide
access to the effects of
frustration in classical and more quantum regimes, as evidenced in
the low-temperature properties
of both the members that possess long-ranged spin structures, and
of those that show more
unconventional orderings (Table \ref{order_type_table}).

\section{The alunite and jarosite families}

In mineralogical terms, the jarosites are a subfamily of the alunite
group. This is a series of compositions
with the formula AB$_3$(SO$_4$)$_2$(OH)$_6$ (where A = Na$^+$, K$^+$,
Ag$^+$, Rb$^+$,
H$_3$O$^+$, NH$_4^+$, $\frac{1}{2}$Ba$^{2+}$, and $\frac{1}{2}$Pb$^{2+}$;
and
B=Al$^{3+}$ and Fe$^{3+}$); alunite itself has the formula
KAl$_3$(SO$_4$)$_2$(OH)$_6$ .\cite{Dana}
While in mineralogy the jarosities are the Fe members of this series, the
term has been expanded by
physicists to include the magnetic members of this series, {\it i.e.}
those where B=Fe$^{3+}$,
Cr$^{3+}$ and V$^{3+}$.\cite{Townsend,Lee_PRB,Wills_Thesis} Thus,
within the jarosites series
the spin states can be tailored by chemical methods. Diamagnetic dilution
studies can also be made
in which a non-magnetic ion, such as Al$^{3+}$, Ga$^{3+}$, or In$^{3+}$,
is substituted onto
the B site.\cite{Dutrizac,Townsend}  In addition to this versitility in
engineering various {\it kagom\'e}
systems, control can also be exercised over the details of the interplane
magnetic exchange by
replacement of the SO$_4^{2-}$ ion by CrO$_4^{2-}$, or
SeO$_4^{2-}$.\cite{Dutrizac}

\begin{figure}

\caption{The crystal structure of jarosite,
KFe$_3$(SO$_4$)$_2$(OH)$_6$.} \label{structure}
\centerline{\hbox{\epsfig{figure=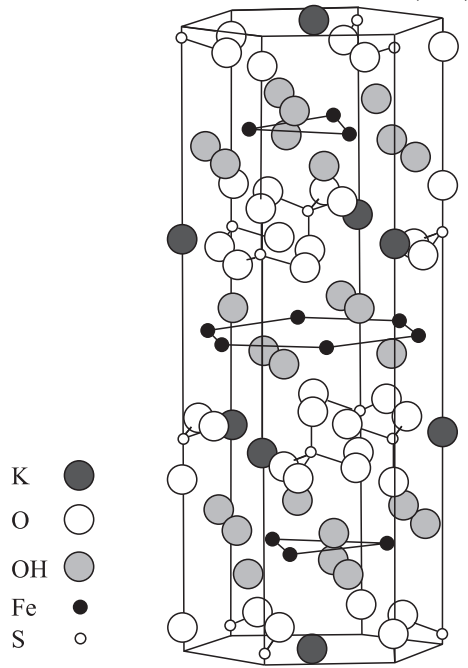,width=6cm}}}
\end{figure}

\section{Crystal structure.}

The crystal structure of the majority of the jarosites is depicted in
Figure \ref{structure}.\cite{Hendricks}
While there has been discussion over whether the actual space group is
$R3$m or $R\bar{3}$m, the
latter is now the generally accepted symmetry. A notable exception to
this is
Pb$_{0.5}$Fe$_3$(SO$_4$)$_2$(OH)$_6$, where segregation of the Pb$^{2+}$
ions into mostly full
and empty layers leads to a doubling of the unit cell along the
{\it c}-direction.\cite{Szymanski} However,
no other deviation away from the rhombohedral jarosite structure is shown
in this material.

The most important features for the magnetic properties are the three
{\it kagom\'e} layers of Fe$^{3+}$
ions with the hexagonal stacking sequence ...ABC... The details of the
intraplane and the interplane
exchange will be examined separately in Sections \ref{Intralayer coupling}
and \ref{Interlayer coupling}.

\section{Fe coordination and exchange pathways}
\label{Fe coordination and exchange pathways}

The magnetic Fe$^{3+}$ ions are coordinated by a distorted octahedron made
up of 4 equatorial
hydroxy and 2 axial sulfate group oxygens, as shown in
Figure \ref{intra_layer_coordination}: selected
bond distances and angles are given in Table \ref{bond_lengths_table}.
The octahedra are tilted with
respect to the crystallographic {\it c}-direction, and it is this canting
that defines the axis of any Ising
single-ion anisotropy for the B$^{3+}$ atoms. XY anisotropy is similarly
defined by the plane of the
4 hydroxy groups.

\subsection{Intralayer coupling}
\label{Intralayer coupling}

\begin{figure}
\caption{The distorted octahedral coordination of the B$^{3+}$
atoms, and their connection via a shared hydroxy oxygen. The third
coordination octahedron has not been highlighted for clarity.
Definitions of atom labels are made for Table
\ref{bond_lengths_table}.}
\centerline{\hbox{\epsfig{figure=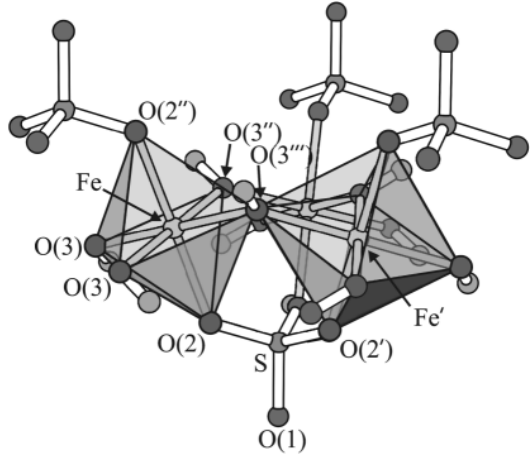,width=7cm}}}
 \label{intra_layer_coordination}
\end{figure}

As shown in Figure \ref{intra_layer_coordination}, simple distance
arguments indicate the strongest
intralayer superexchange between nearest-neighbour B atoms is a pairwise
interaction that takes place
via a shared hydroxide group with a bridging angle of close to 133$^\circ$.
Unfortunately, the
complexity of the superexchange at a non-linear or non-orthogonal angle
prevents an explanation
from being made of the exchange observed in these systems.

Also possible is exchange via the sulfate group (labelled S in the Figure).
This is likely to be far weaker
than that mediated by the bridging hydroxy group because it involves a greater
number of chemical bonds.
In the jarosite structure this sulfate is shared in the same way between
the three B octahedra of a
{\it kagom\'e} triangle, and so it will act to equally couple the three magnetic
atoms of the triangle.

\subsection{Interlayer coupling}
\label{Interlayer coupling}

\begin{figure}

\caption{Exchange paths between the {\it kagom\'e} layers in the
jarosite structure.} \label{interlayer exchange}
\centerline{\hbox{\epsfig{figure=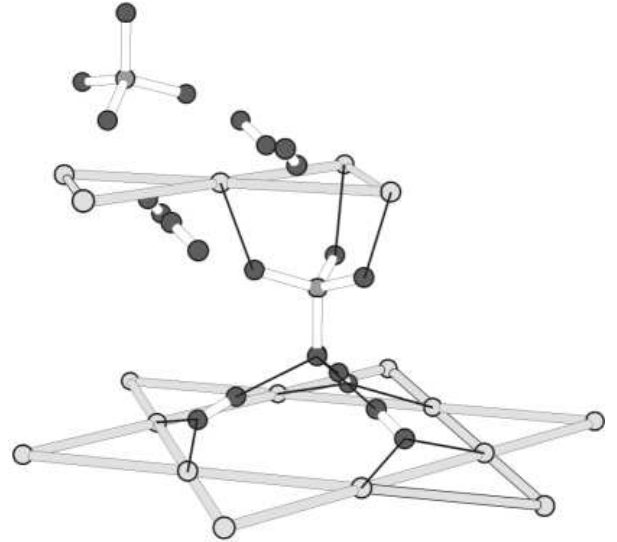,width=8cm}}}
\end{figure}

The expectation that the jarosites could be tailored into providing good
model {\it kagom\'e} systems
comes from the large separation of the {\it kagom\'e} planes, with respect to
those within the plane: in (H$_3$O)Fe$_3$(SO$_4$)$_2$(OH)$_6$ the nearest
neighbour B---B distance 3.66 \AA,
whereas the distance between {\it kagom\'e} layers is  5.94 \AA. The exchange
between the {\it kagom\'e} layers
is therefore anticipated to be far weaker than that within a layer from
simple arguments of distance
and the number of chemical bonds involved in the superexchange
(Figure \ref{interlayer exchange}) .
Despite its relative weakness, it  is apparent from the different propagation
vectors observed in KFe$_3$(SO$_4$)$_2$(OH)$_6$ and
KFe$_3$(CrO$_4$)$_2$(OH)$_6$
that the interlayer
coupling involving the XO$_4$ ion plays a definitive r\^ole in the observed
magnetic structure.

In a recent work, symmetry analysis has been used to determine the stability
conditions of the
different magnetic structures observed in the jarosites.\cite{Wills_inprep}
In terms of an exchange
constant relating the nearest B$^{3+}$ atoms of neighbouring layers, it has been
found that the different
${\bf k}=00\frac{3}{2}$ spin configurations observed in
KFe$_3$(SO$_4$)$_2$(OD)$_6$\cite{Inami_PRB,Frunzke}
and (H$_3$O)V$_3$(SO$_4$)$_2$(OH)$_6$\cite{Wills_inprep} are stabilised by ferromagnetic and
antiferromagnetic
exchange respectively, while the ${\bf k}=000$ spin structure observed in  both
KCr$_3$(SO$_4$)$_2$(OD)$_6$\cite{Lee_PRB}
and KFe$_3$(CrO$_4$)$_2$(OD)$_6$\cite{Townsend} is stabilised by antiferromagnetic
coupling.

\begin{figure}

\caption{The linkage of the distorted coordination octahedra
around the B$^{3+}$ to form the {\it kagom\'e} lattice.}
\centerline{\hbox{\epsfig{figure=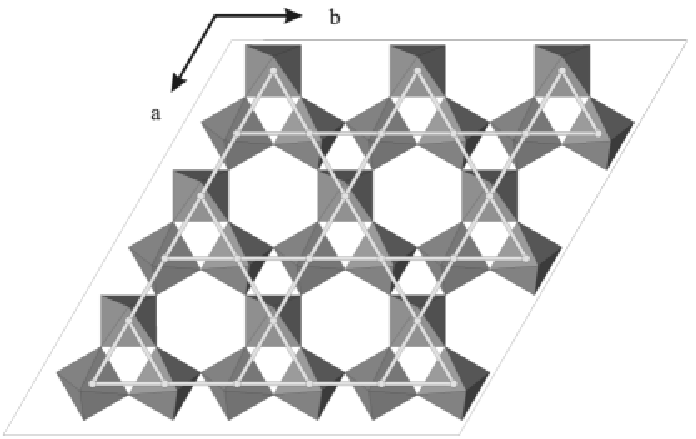,width=8cm}}}
\label{kagome
lattice}
\end{figure}

\subsection{Effects of A cation}

 The marked difference in the responses of hydronium jarosite from the
other iron jarosites\cite{Wills_Faraday,Wills_EPL,Wills_PRB} indicates
that the A cation is capable of having a profound influence on the
low-temperature magnetic properties
of these systems. This is most likely a consequence of the non-spherical
symmetry of the H$_3$O$^+$
ion and its ability to hydrogen bond with some of the 12 sulfate group
oxygens that coordinate the
A site. When the H$_3$O ions are no longer orientationally mobile, as
is the situation at low
temperature, these hydrogen bonds will create interactions between
neighbouring sulfate groups that are
not present in the other jarosites. Due to the complexity this would
generate, it is not yet clear if they
will act to hinder or improve the interlayer exchange. However, the
large difference in the magnetic
properties of the hydronium jarosite from those of the monatomic ions,
even for compositions with the
same Fe atom occupation,\cite{Frunzke} suggests that there is indeed a disruption of
simple interlayer exchange
couplings, and concomitantly a destabilisation of any long-range ordered
state that they would induce.

That the ammonium salt\cite{Wills_PRB} behaves like the other
spherical A-site metals, suggests that its
hydrogen bonding is either less disturbing, or of a lesser consequence,
than that of H$_3$O. Further
studies are required before the effects of the A cation can be understood.

\section{Influences of nonstoichiometry.}

It is well known that the jarosites are subject to
nonstoichiometry on both the A and the B sites. Deficiency of the
A metal or ammonium cation is charge compensated for by
incorporation of H$_3$O$^+$ ions onto the A site, so that the site
occupation remains essentially at unity.\cite{Kubisz} The extreme
limit of this situation is of course the hydronium salt itself.
Under-occupation of the B site results in protonation of the
hydroxy groups in the structure. As these groups play the dominant
r\^ole in the mediation of the superexchange between the metal
centres, it is conceivable that such disorder and the associated
randomness in the exchange interactions, could have important
consequences for the magnetic properties.

\section{Conventional Orderings}

Long-range magnetic structures with the propagation
vectors\footnote{both propagation vectors and Wyckoff labels refer
to the non-primitive hexagonal setting of $R\bar{3}$m } ${\bf
k}=000$ and ${\bf k}=00\frac{3}{2}$ have been observed in the
jarosites (see Table \ref{order_type_table}). Representational
Analysis calculations have shown that there are only 3 non-zero
irreducible representations for the 9d B$^{3+}$ site and that the
basis vectors associated with them are identical  for both values
of the propagation vector; they are shown in Figure \ref{Basis
vectors}.\cite{Wills_inprep} These calculations show that for both
of these values of {\bf k}, the same inplane spin configurations
are allowed, and that the relation between moments in successive
{\it kagom\'e} layers is determined simply by the value of {\bf
k}: in the case of {\bf k}=0, the moments related by unit
translations of the primitive cell are ferromagnetically aligned,
while those for ${\bf k}=00 \frac{3}{2}$ are antiferromagnetically
aligned.

The basis vectors for the unrepeated first order irreducible representation,
here labelled $\Gamma_1$,
correspond to a 120$^\circ$ structure in which the moments are fixed
along particular crystallographic axes.
Those of the doubly-repeated irreducible representation,
$\Gamma_3$, form an umbrella structure
made up of a 120$^\circ$ configuration, in which the in-plane component
of the spins are related by
60$^\circ$ to those of $\Gamma_1$, and a ferromagnetic component along
the crystallographic {\it c}- axis.

It is notable that both these 120$^\circ$ structures possess a uniform
chirality, \boldmath $\kappa$ \unboldmath $= +1$.
Where \boldmath $\kappa$ \unboldmath is defined as the pairwise vector product clockwise around
a triangle:

\begin{equation}
\textrm{\boldmath $\kappa$ \unboldmath} =\frac{2}{\left(3\sqrt{3}\right)}\left[{\bf S}_{1}\times
{\bf S}_{2}+{\bf S}_{2}\times {\bf S}_{3} + {\bf S}_{3}\times {\bf
S}_{1}\right] ,
\label{chirality}
\end{equation}

\noindent
Thus, it is a consequence of symmetry, rather than
anisotropy\cite{Inami_PRB}, that the $q=0$
spin configurations with \boldmath $\kappa$ \unboldmath $= -1$ are forbidden.

The third irreducible representation, $\Gamma_6$, is 2 dimensional and is
repeated 3 times.
This  leads to a total of 6 basis vectors which describe in general a
complex spin configuration. However,
conjugate pairing on these basis vectors leads to only 3 being required
to describe all the possible orderings
under this irreducible representation. Notably, coplanar ordering under
this representation involves
non-spin compensated triangles, {\it i.e.}

\begin{equation}
\sum_i {\mathbf S}\not=0
\label{Eq1}
\end{equation}

\noindent
Where the sum is over the moments of an individual triangle. There is
therefore a net moment associated
with each of the {\it kagom\'e} planes.

\begin{figure}
\centerline{\hbox{\epsfig{figure=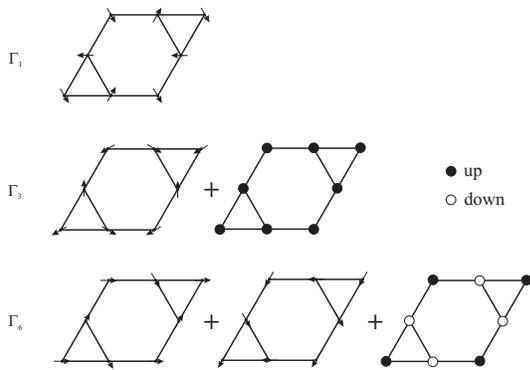,width=7cm}}}
\caption{The motifs for the different basis vectors for the
non-zero irreducible representations of the point group
$D^{5}_{3d}$ at the B site for the propagation vectors {\bf k}=0
and {\bf k}$=0  0  \frac{3}{2}$. The labelling scheme follows that
for {\bf k}=0.\cite{Wills_inprep}} \label{Basis vectors}
\end{figure}

\subsection{${\bf k}=000$}

Magnetic ordering with the magnetic cell the same size as the nuclear
has been observed in the
compositions KFe$_3$(CrO$_4$)$_2$(OD)$_6$\cite{Townsend} and
KCr$_3$(SO$_4$)$_2$(OD)$_6$.\cite{Townsend,Lee_PRB} In both cases it is believed that the
nearest neighbour antiferromagnetic exchange results in a 120$^\circ$ spin
structure, while some
secondary interaction such a single-ion anisotropy leads to a small canting
of the moments along
the {\it c}-axis. The presence of such a ferromagnetic component in a system
with antiferromagnetic
nearest-neighbour interactions is compatible only with ordering under
the repeated first-order
representation, $\Gamma_3$. The general structure may thus be described
by an umbrella mode
involving some linear combination of the two basis vectors.

\subsection{${\bf k}=00\frac{3}{2}$}

Ordering with this propagation vector is found in the majority of the
iron jarosites
(AFe$_3$(SO$_4$)$_2$(OD)$_6$, where A = Na$^+$, K$^+$, Ag$^+$, Rb$^+$,
Tl$^+$, H$_3$O$^+$, and ND$_4^+$)  and the vanadium salt
(H$_3$O)V$_3$(SO$_4$)$_2$(OH)$_6$. In all the Fe compositions,
ordering under the repeated first-order representation is again
observed.\cite{Wills_inprep}
Hydronium vanadium jarosite in contrast, shows ordering under the
representation
$\Gamma_6$.\cite{Wills_inprep}

At present there are still questions over the details of the
ordering transitions in the iron jarosites as some samples exhibit
two ordering transitions\cite{Maegawa_JPSJpn,Wills_PRB,Frunzke}
while others display only one.\cite{Inami_PRB} Recent refinement
of neutron powder diffraction data collected from
KFe$_3$(SO$_4$)$_2$(OD)$_6$ have shown that in the samples with
two transitions the intermediate phase is an umbrella structure
based on $\Gamma_3$. Muon Spin Relaxation (MuSR) experiments
revealed the presence of large fluctuations in this intermediate
structure that reduce the size of the ordered moment. Harrison
{\it et al.}\cite{Cryopad} confirmed the absence of an
out-of-plane component to the ordering by single crystal neutron
diffraction from a natural sample using the technique of spherical
polarisation analysis, and demonstrated by powder neutron
diffraction that the lower-temperature transition, in which the
moments drop into the {\it kagom\'e} plane, is disrupted by
disorder on the magnetic B sites.\cite{Frunzke} The display of a
single transition is therefore likely to be consequence of sample
nonstoichiometry. As the out-of-plane component of the
intermediate phase is not stabilised by exchange interactions, it
suggests the presence of some other influence, such as Ising
single-ion anisotropy. It is however difficult to imagine that
this anisotropy changes to being XY so close to the first ordering
temperature (in KFe$_3$(SO$_4$)$_2$(OD)$_6$, $T_{C1}$=64.5 K and
$T_{C2}$=57 K), and the reasons for the second transition are
consequentially still unclear.

The magnetic structure observed in  (H$_3$O)V$_3$(SO$_4$)$_2$(OH)$_6$ will
be discussed later in
Section \ref{S=1 Unconventional long-range order} as its origins appear
to be unconventional.

\section{Unconventional Orderings.}
This section presents a brief review of the unconventional orderings
observed in the jarosites as a function
of the different spin values.

\subsection{$S=\frac{5}{2}$ Topological Spin Glass.}

The most studied member of the jarosite series is the hydronium
salt (H$_3$O)Fe$_3$(SO$_4$)$_2$(OH)$_6$, which is an example of an
unconventional spin
glass.\cite{Wills_Faraday,Wills_EPL,Wills_PRB,Wills_CondMat}
Before describing its responses, it is useful to examine the
occurence of glassy magnetic phases in geometrically frustrated
antiferromagnets. In relation to the underconstrained lattices,
this possibility was first studied by Villain\cite{Villain}, who
noted that the high degeneracy of the magnetic {\it pyrochlore}
sublattice of the B-site of the spinels  is robust to the
introduction of a small degree of disorder. This conclusion can be
trivially extended to the {\it kagom\'e} antiferromaget, and so
establishes that small amounts of disorder should not give rise to
a spin glass phase. Villain further noted that if the degree of
disorder was sufficient to break the degeneracies present in the
system, then the spin glass-like state that would result would not
necessarily be typical as the large degree of frustration in these
systems could still lead to unconventional physics. There is
therefore a question that hangs over any unconventional spin glass
phases observed in the highly frustrated lattices: {\it are they
the result of disorder in the sample, whether related to sites or
bonding (exchange) ?}

With an occupation of the magnetic sublattice of $\sim 97$ \%, the degree
of site-disorder present in
hydronium jarosite is not expected to be at the level required to stabilise
spin glass behaviour.\cite{Wills_Faraday}
The vitreous magnetic phase observed at low temperature is ergo quite
remarkable. Measurements of the
thermodynamic and kinetic properties confirm its notability as they
indicate clearly that the glassy magnetic
phase seen below a critical transition at $T_g \sim 17$ K\cite{Wills_CondMat}
is unlike those of conventional
site-disordered spin glasses\cite{Mydosh}: the magnetic contribution to the
specific heat follows a $T^2$
relation with temperature,\cite{Wills_EPL} as opposed to the linear law
typical of site-disordered systems,
and the temperature dependence of the out-of-equilibrium dynamics is remarkably
weak.\cite{Wills_CondMat} The latter provides interesting information
about the spin glass state itself: if
the relaxation of a spin glass is envisaged as involving the movement
towards some equilibrium value, these
results indicates that this equilibrium state changes {\it far} slower with temperature
in hydronium jarosite than in conventional
spin glass systems, a situation that is intriguing reminiscent of a system
moving  through a highly degenerate
ground state manifold.

Polarised neutron powder diffraction studies of
(D$_3$O)Fe$_3$(SO$_4$)$_2$(OD)$_6$ have shown that despite a
freezing temperature of $T_g \sim 17$ K, short-ranged correlations
are present at temperatures as high as 250 K, demonstrating well
the difficulty with which the moments are
freezing.\cite{Oakley,Wills_D7} The depression of the freezing
temperature with relation to the exchange energy is compatible
with suggestions of the freezing being a consequence of a new
energy scale--- a reduced Kosterlitz-Thouless temperature--- that
approximates to $\theta/48$ for an XY
system.\cite{Spin_fold_1,Spin_fold_2} The magnitude and spatial
extension of the correlations increase only gradually with
cooling, until they reach an apparent saturation value of $\sim
10$ \AA  ~ at 1.5 K.

Another unorthodox feature of this unconventional spin glass state is
its fragility. In sharp contrast with
site-disordered spin glasses, where an increase in disorder has only
a small effect on the magnetic properties,
in hydronium jarosite a deliberate increase in the level of disorder
to $\sim 10$ \% is found to induce
long-range magnetic order.\cite{Wills_PRB} This is an
example of  `order-by-disorder',\cite{Villain_OBD,Reimers,Chalker}
where increased disorder acts to stabilise the formation of
long-range N\'eel order with respect to a spin glass state.

While the question of the influences of disorder remains unanswered in
(H$_3$O)Fe$_3$(SO$_4$)$_2$(OH)$_6$,
the unconventional nature of the observed spin glass phase is manifest.
It is clear that the origins of the spin glass
state itself and these unconventional properties are related to the
frustrated topology of the {\it kagom\'e} lattice. This
has lead to hydronium jarosite being termed a {\it topological spin
glass}.\cite{Wills_CondMat}

\subsection{$S=\frac{3}{2}$ Highly Fluctuating Ground States.}

The apparent nonconformity of the magnetic properties of the hydronium
jarosites when compared with the
other Fe jarosites holds true also for the chromium analogue. While
KCr$_3$(SO$_4$)$_2$(OD)$_6$
shows a transition to long-range order\cite{Lee_PRB}, albeit with a
greatly reduced sublattice magnetisation,
due to the presence of quantum fluctuations, hydronium chromium
jarosite (H$_3$O)Cr$_3$(SO$_4$)$_2$(OH)$_6$
displays quite different behaviour.\cite{Wills_Thesis} Despite strong
antiferromagnetic exchange
($\theta\simeq -60$ K), at low temperatures only a broad feature
corresponding to the onset of short-ranged
correlations is observed at $\sim 30$ K in the dc susceptibility,
before a small ferromagnetic transition at 2.2 K.
The ferromagnetic nature of the correlations involved in this transition
is confirmed by the field-dependence of the
specific heat data below $\sim 5$ K. The magnetic entropy associated
with the low temperature transition is
only 5.4 \% is the value expected for a $S=\frac{3}{2}$ system and
suggests the presence of extensive quantum
fluctuations.

\subsection{$S=1$ Unconventional long-range order.}
\label{S=1 Unconventional long-range order}

The $S=1$ {\it kagom\'e} system (H$_3$O)V$_3$(SO$_4$)$_2$(OH)$_6$
has been shown to possess long-range order below a critical
transition of $T_C \simeq 20$ K.\cite{Wills_Thesis,Wills_inprep}
However, the magnetic structure observed is remarkable because
despite the non-zero exchange field experienced by each moment,
some are orientated such as to nullify their exchange
energy.\cite{Wills_inprep} It therefore appears that simple
arguments of the long-range order being the result of minimised
exchange energy cannot explain the formation of this ground state
configuration. Rather, recent work has suggested that it is the
result of a particular combination of ferromagnetic
nearest-neighbour exchange and Ising anisotropies. Notably, these
are precisely the types required to stabilise a spin ice phase on
the kagom\'e lattice.\cite{kagome spin ice}

the additional degeneracies that are afforded by these particular
spin orientations appear to be the cause of their stabilisation.

\section{Discussion and Conclusion}

There are still numerous questions concerning the effects of nonstoichiometry
and how substitution of the different
ions influences the magnetic exchange in the jarosites. In
particular, the reasons for the apparent distinction of
the hydronium salts remains unclear. Unfortunately, these questions
are difficult to answer experimentally
because synthetic limitations require that the majority of experiments
are carried out on powders. Inferences
made from the materials that show more comprehensible long-range
orderings are consequently important
in aiding understanding of those that show more unconventional order.
This is well exemplified by the
apparent planar anisotropy at low temperature in
KFe$_3$(SO$_4$)$_2$(OD)$_6$. Its observation is
important, as such an anisotropy is required by a theoretical model of an
unconventional spin glass ground
state on the {\it kagom\'e} antiferromagnet.\cite{Spin_fold_1,Spin_fold_2}

As this paper has shown, the members of the jarosite family
display a wide variety of interesting physics. The unconventional
spin glass, highly fluctuating, and unconventional long-range
magnetic orderings show well different influences of frustration
that are still far from being explained. It is hoped that this
review will incite further work on these intriguing materials.


\newpage

\begin{table}
\caption{Values of the intercept of the inverse susceptibility
with the temperature axis $\theta$, the exchange constant $J$, the
critical temperature $T_C$, and order types for various members of
the family AB$_3$(XO$_4$)$_2$(OH)$_6$. As series expansion
calculations\cite{Harris,Reimers_PRB_1992} indicate that a
modified form of the Curie-Weiss law holds below $T\simeq
JS(S+1)/k$, $J$ is calculated from $\theta=-2JS(S+1)/k_B$ for the
Fe jarosites, while the conventional mean field result
$\theta=-(4/3k_B)JS(S+1)$ is used for the other compositions.}
\begin{tabular}{cccccc}
Formula  & $\theta$ /K & $J$ /K& $T_C$ /K & Ordering type & Ref. \\
\tableline
NaFe$_3$(SO$_4$)$_2$(OH)$_6$& -730& 41.7& 60, 54 & ${\bf k}=0 0 \frac{3}{2}$  & \cite{Maegawa_JPSJpn} \\
NaFe$_3$(SO$_4$)$_2$(OH)$_6$& --667&38.1  & 62, 42 &${\bf k}=0 0 \frac{3}{2}$  & \cite{Wills_PRB} \\
KFe$_3$(SO$_4$)$_2$(OH)$_6$& -700& 40.0 & 65&${\bf k}=0 0 \frac{3}{2}$  & \cite{Inami_PRB} \\
KFe$_3$(SO$_4$)$_2$(OH)$_6$& -663 & 37.9& 64.5, 57.0&${\bf k}=0 0 \frac{3}{2}$  & \cite{Frunzke} \\
(NH$_4$)Fe$_3$(SO$_4$)$_2$(OH)$_6$ & -670& 38.3 & 61.7, 57.1& ${\bf k}=0 0 \frac{3}{2}$  & \cite{Maegawa_JPSJpn} \\
(ND$_4$)Fe$_3$(SO$_4$)$_2$(OD)$_6$  & -640 & 36.6& 62, 46& ${\bf k}=0 0 \frac{3}{2}$  & \cite{Wills_PRB}  \\
Pb$_{0.5}$Fe$_3$(SO$_4$)$_2$(OH)$_6$& - &   - & -& ${\bf k}=0 0 \frac{3}{2}$  & \cite{Collins_unpublished}\\
TlFe$_3$(SO$_4$)$_2$(OH)$_6$ &  - &  - & - & ${\bf k}=0 0 \frac{3}{2}$  & \cite{Collins_unpublished}\\
AgFe$_3$(SO$_4$)$_2$(OD)$_6$ & -677 & 38.7 & 51& ${\bf k}=0 0 \frac{3}{2}$  & \cite{Wills_PRB}  \\
RbFe$_3$(SO$_4$)$_2$(OD)$_6$ & -688 & 39.3 & 47& ${\bf k}=0 0 \frac{3}{2}$  & \cite{Wills_PRB}  \\
(D$_3$O)Fe$_{3-x}$Al$_y$(SO$_4$)$_2$(OD)$_6$ & -720& 41.1 & 41.1& ${\bf k}=0 0 \frac{3}{2}$  & \cite{Wills_PRB}  \\
(D$_3$O)Fe$_3$(SO$_4$)$_2$(OD)$_6$   & -700 & 40.0& 13.8 & \stk{unconventional}{spin glass} & \cite{Wills_Faraday,Wills_EPL,Wills_CondMat} \\
 & & & & & \\
KFe$_3$(CrO$_4$)$_2$(OH)$_6$ & -600 & 34.2& 65& ${\bf k}=0 0 0$   & \cite{Townsend} \\
 & & & & & \\
KCr$_3$(SO$_4$)$_2$(OH)$_6$   & -70& 14 & 1.6 & ${\bf k}=0 0 0$ & \cite{Lee_PRB}\\
(H$_3$O)Cr$_3$(SO$_4$)$_2$(OD)$_6$   & -78 & 15.6& 1.2  & unknown    & \cite{Wills_Thesis} \\
 & & & & & \\
(H$_3$O)V$_3$(SO$_4$)$_2$(OD)$_6$ & -45& 9 & 21 & ${\bf k}=0 0 \frac{3}{2}$   & \cite{Wills_inprep} \\
\end{tabular}
\label{order_type_table}
\end{table}
\begin{table}
\caption{Selected bond lengths and angles for
(H$_3$O)Fe$_3$(SO$_4$)$_2$(OH)$_6$ at 1.5 K.\cite{Wills_Faraday}}
\begin{tabular}{ccc}
Bond  & Bond length (\AA)& Bond Angle ($^\circ$) \\
\tableline
Fe---Fe    & 3.66228 (6)  & \\
Fe---O(2) & 2.0374 (19) & \\
Fe---O(3) & 1.9921(8)  & \\
S---O(1) & 1.492 (5)  & \\
S---O(2) & 1.4495 (26)  \\
O(2)---O(3) & 2.8236 (15) & \\
O($2''$)---O(3) & 2.8750 (27)  & \\
O(3)---O($3'$) & 2.8103 (33)  & \\
O($3'$)---O($3''$) & 2.8240 (20)  & \\
O(2)---S---O($2'$) &  & 110.83 (23)   \\
O(2)---S---O(1) & & 108.08 (24) \\
S---O(2)---Fe & & 129.265 (8) \\
Fe---O(3)---Fe$'$ &  & 133.625 (1)\\
O(2)---Fe---O(3) &  & 88.968 (13) \\
O(3)---Fe---O($3'$) &  & 89.721 (13) \\
O($3'$)---Fe---O($3''$) & & 90.279 (19) \\
\end{tabular}
\label{bond_lengths_table}
\end{table}


\begin{references}

\bibitem[*]{ASW} Present address: Institut Laue-Langevin, 6 rue Jules
Horowitz, BP 156, 38042 Grenoble Cedex 9, France.

\bibitem{Harris}
A.~B. Harris, C. Kallin, and A.~J. Berlinsky, Phys. Rev. B {\bf
45}, 2899 (1992).

\bibitem{Reimers_PRB_1992}
J.~N. Reimers and A.~J. Berlinsky, Phys. Rev. B {\bf 48}, 9539
(1993).


\bibitem{Maegawa_JPSJpn}
S. Maegawa, M. Nishiyama, N. Tanaka, A. Oyamada, and M. Takano,
J. Phys. Soc. Jpn. {\bf 65}, 2776 (1996).

\bibitem{Wills_PRB}
A.~S. Wills, A. Harrison, C. Ritter, and R.~I. Smith, Phys. Rev. B
{\bf 61}, 6156 (2000).

\bibitem{Inami_PRB}
T. Inami, M. Nishiyama, S. Maegawa, and Y. Oka, Phys. Rev. B {\bf 61}, 12181 (2000).

\bibitem{Frunzke}
J. Frunzke, T. Hansen, A. Harrison, J.~S. Lord, G.~S. Oakley, D.
Visser, and A.~S. Wills, J. Mater. Chem. {\it In press} (2001).


\bibitem{Collins_unpublished}
M.~F. Collins, {\it Unpublished work}.

\bibitem{Wills_Faraday}
A.~S. Wills and A. Harrison, J. Chem. Soc. Faraday Trans. {\bf
92}, 2161 (1996).

\bibitem{Wills_EPL}
A.~S. Wills, A. Harrison, S.~A.~M. Mentink, T.~E. Mason, and Z.
Tun, Europhys. Lett. {\bf 42}, 325 (1998).

\bibitem{Wills_CondMat}
A.~S. Wills, V. Depuis, E. Vincent and R. Calemczuk, Phys. Rev. B
{\bf   62}, R9264 (2000).


\bibitem{Townsend}
M.~G. Townsend, G. Longworth, and E. Roudaut, Phys. Rev. B {\bf
33}, 4919 (1986).

\bibitem{Lee_PRB}
S.-H. Lee, C. Broholm, M.~F. Collins, L. Heller, A.~P. Ramirez,
Ch. Kloc, E. Bucher, R.~W. Erwin, and N. Lacevic, Phys. Rev. B
{\bf 56}, 8091 (1997).

\bibitem{Wills_Thesis}
A.~S. Wills, {\it Ph.D. Thesis}, The University of Edinburgh
(1997)


\bibitem{Wills_inprep}
A.~S. Wills, Phys. Rev. B, {\bf 63}, 064430 (2001).


\bibitem{Dana}
C. Palache, H. Berman, and C. Frondel, {\it Dana's system of mineralogy}
(John Wiley and Sons, London, 1951).

\bibitem{Dutrizac}
J.~E. Dutrizac, NATO Conf. Ser. (Hydrometall. Process. Fund.)
(1984).

\bibitem{Hendricks}
S.~B. Hendricks, Am. Miner. {\bf 22}, 773 (1937).

\bibitem{Szymanski}
J.~T. Szymanski, Can. Miner. {\bf 23}, 659 (1985).

\bibitem{Kubisz}
J. Kubisz, Miner. Pol. {\bf 1}, 45 (1970).

\bibitem{Cryopad}
E. Lelievre-Berna, A. Harrison,  G. Oakley, and D. Visser, Annual
Report of the Institute Laue Langevin for 1999 (2000) expt 5-61-37.

\bibitem{Villain}
J. Villain, J. Phys. (Paris) {\bf 38}, 26 (1977).

\bibitem{Mydosh}
J.~A. Mydosh, {\it Spin glasses: an experimental introduction}
(Taylor \& Francis Ltd, London 1993).

\bibitem{Oakley}
G.~S. Oakley, D. Visser, J. Frunzke, K.~H. Andersen, A.~S. Wills,
and A. Harrison, Physica B {\bf 267-268}, 142 (1999).

\bibitem{Wills_D7}
A.~S. Wills, G.~S. Oakley, D. Visser, J. Frunzke, A. Harrison, and
K.~H. Andersen, Phys. Rev. B {\it in press} (2001).

\bibitem{Spin_fold_1}
I. Ritchey, P. Chandra, and P. Coleman, Phys. Rev. B {\bf 47}, 15342 (1993).

\bibitem{Spin_fold_2}
P. Chandra, P. Coleman, and I. Ritchey, J. Physique I {\bf 3}, 591 (1993).

\bibitem{Villain_OBD}
J. Villain, R. Bidaux, J.-P. Carton, and R. Conte, J. Physique
{\bf 41}, 1263 (1980).

\bibitem{Reimers}
J.~N. Reimers, Phys. Rev. {\bf  B 45}, 7287 (1992).

\bibitem{Chalker}
J.~T. Chalker, P.~C.~W. Holdsworth, and E.~F. Shender, Phys. Rev.
Lett. {\bf 68}, 855 (1992).

\bibitem{kagome spin ice}
A.~S. Wills, R. Ballou, and C. Larcoix, {\it paper in preparation}
(2001).




\end{references}
\end{document}